\documentclass[12pt]{iopart}
\usepackage{iopams}  
\usepackage{graphicx}
\usepackage{lineno}
\usepackage{wrapfig}
\usepackage{tablefootnote}
\usepackage[breaklinks=true,colorlinks=true,linkcolor=blue,urlcolor=blue,citecolor=blue]{hyperref}

\def\ioptwocol{\setlength\hoffset{-0.5in}\setlength\voffset{-0.5in}\setlength\textwidth{6.75in}
\setlength\columnsep{0.2in}\setlength\textheight{9.25in}\mathindent=0in\twocolumn}

\begin{document}

\title[Biorthogonal projected energies of a Gutzwiller similarity transformed Hamiltonian]{Biorthogonal projected energies of a Gutzwiller similarity transformed Hamiltonian}

\author{J M Wahlen-Strothman$^1$, G E Scuseria$^{1,2,3}$}
\address{$^1$Department of Physics and Astronomy, Rice University, Houston, USA 77005}
\address{$^2$Department of Chemistry, Rice University, Houston, USA 77005}
\address{$^3$Department of Materials Science and NanoEngineering, Rice University, Houston, USA 77005}
\eads{\mailto{guscus@rice.edu}}

\begin{abstract}
We present a method incorporating biorthogonal orbital-optimization, symmetry projection, and double-occupancy screening with a non-unitary similarity transformation generated by the Gutzwiller factor $ n_{i\uparrow}n_{i\downarrow}$, and apply it to the Hubbard model. Energies are calculated with mean-field computational scaling with high-quality results comparable to coupled cluster singles and doubles. This builds on previous work performing similarity transformations with more general, two-body Jastrow-style correlators. The theory is tested on two-dimensional lattices ranging from small systems into the thermodynamic limit and is compared to available reference data.
\end{abstract}

\vspace{2pc}
\noindent{\it Keywords}: Gutzwiller similarity transformation, Hubbard model, symmetry projection

\ioptwocol

\section{Introduction}

Similarity transformations provide a convenient tool to treat many-body correlated effects with wavefunction methods leading to powerful approximations with simple expressions and calculations. These are canonical transformations that preserve the spectrum of the Hamiltonian while casting it in a new basis. Real and Hilbert space Jastrow factors are a popular alternative to the particle-hole excitation operators used in methods such as coupled-cluster theory~\cite{ref1} and have been applied as correlation factors in wavefunctions for Monte Carlo calculations and as similarity transformations in transcorrelation and other methods.~\cite{ref2,ref3,ref4,ref5,ref6,ref7}

In a previous paper,~\cite{ref8} we presented a transformation with two-body, Hilbert space, Jastrow-type operators $J$ that improved a trial Slater determinant wavefunction by introducing short and long-range correlations. Ideally the correlation factors would be applied directly to a simple wavefunction,
\begin{center}\begin{equation}
|J\rangle \propto e^{J}|\Phi\rangle,
\end{equation}\end{center}
but evaluating an expectation value with the correlated wavefunction $|J\rangle$ is generally intractable without using Monte Carlo sampling. As an alternative, we apply the correlation factor as a similarity transformation to the Hamiltonian.~\cite{ref1,ref3,ref8} If~$|E\rangle$ is an eigenstate of the Hamiltonian, we can express it in terms of a new wavefunction $|\Psi\rangle$ as,
\begin{equation}
|E\rangle = e^J|\Psi\rangle.
\end{equation}
We rewrite Schr\"{o}dinger's equation for the wavefunction $|\Psi\rangle$ in terms of a transformed Hamiltonian operator,
\begin{eqnarray}
 &H  e^{J} & |\Psi\rangle = Ee^J|\Psi\rangle, \nonumber \\
  e^{-J} & H  e^J & |\Psi\rangle = E|\Psi\rangle, \nonumber \\
 &\hspace{2ex}\overline H & |\Psi\rangle = E|\Psi\rangle.\label{eq3}
\end{eqnarray}
We aim to find approximate solutions to Schrodinger's equation for this new Hamiltonian rather than attempt to directly evaluate expectation values with a correlation factor acting on a mean-field wavefunction. In this method, we approximate $|\Psi\rangle$ as a single Slater determinant or symmetry projected Slater determinant wavefunction.~\cite{ref10} We then build a system of equations in order to calculate the single particle basis of the Slater determinant and the correlation parameters in $J$ such that Eq.~(\ref{eq3}) is satisfied within a subspace of the full Hilbert space.

Here, we extend our consideration of transformations with Jastrow-type operators from previous work.~\cite{ref8} In this case, we limit the correlations to short range interactions from the Gutzwiller factor $n_{i\uparrow}n_{i\downarrow}$. This transformation was previously applied~\cite{ref5} as a Hilbert space form of transcorrelation and optimized by minimizing the energy variance for some small Hubbard chains. This correlation factor has also been used in variational methods such as Monte Carlo~\cite{ref2} and as a projection operator for resonating valence bond states.~\cite{ref9} While removal of the longer range correlation factors reduces the overall flexibility of the ansatz, the short range properties focused on in this work are still treated accurately and are most important for systems with short range interaction, such as the Hubbard model.~\cite{ref8} The transformation produces a non-Hermitian, many-body Hamiltonian that can be easily evaluated in mean-field with low computational scaling. The correlation amplitudes satisfy a set of projected equations similar to coupled cluster theory, and the simplicity of the resulting equations compared to other transformation methods leads readily to further improvements in the method and application of more advanced reference wavefunctions.

\section{Theory}

In order to obtain accurate results, we need to start with a reference wavefunction that requires relatively small corrections from the correlation factors. We will use a spin-projected reference wavefunction as it is reasonably accurate in both strongly and weakly correlated regimes.

\subsection{Spin Projection}

Restricted Hartree-Fock (RHF) is a common mean-field approximation consisting of a single Slater determinant wavefunction that preserves the spin symmetries of the Hamiltonian. However, this approximation is very poor for strongly correlated systems. For example, it cannot form the N\'{e}el state required to describe many spin lattice systems. Unrestricted Hartree-Fock (UHF) preserves $S_z$ symmetry but breaks $S^2$ symmetry by allowing different configurations of the up and down spins. This is known as the collinear case as opposed to the non-collinear case, referred to as generalized Hartree-Fock (GHF), where the up and down spins are mixed.  UHF can significantly improve energies in strongly correlated systems, but other properties, such as spin correlations, may be poorly described.  In order to incorporate this improvement without sacrificing the symmetries and quality of the wavefunction we use the spin-projected unrestricted Hartree-Fock (SUHF) ansatz,~\cite{ref10}
\begin{center}\begin{equation}
|SUHF\rangle \propto P^s_{mm} |\Phi\rangle,
\end{equation}\end{center}
where 
$|\Phi\rangle$ 
is a UHF state. Any spin contamination is projected out by the operator $P$, resulting in a wavefunction with proper symmetry that is not a single Slater determinant. The operator used in this work is,~\cite{ref10}
\begin{eqnarray}
P^s_{mk} &=& |s;m\rangle\langle s;k| \nonumber\\
P^s_{mk} &=& \frac{2s+1}{8\pi^2}\int \mathrm d\Omega D^{s*}_{mk}(\Omega)R(\Omega), \label{eq4}
\end{eqnarray}
where $R(\Omega) = e^{\mathrm{i}\alpha S_z}e^{\mathrm{i}\beta S_y}e^{\mathrm{i}\gamma S_z}$ and $D^s_{mk}(\Omega) = \langle s;m|R(\Omega)|s;k\rangle$ is the Wigner D-matrix. This operator projects a broken symmetry state into a wavefunction with spin quantum numbers $S$ and $m$ by enforcing that the state is invariant to spin rotations. As the projection is an integration over a manifold of spin rotated states, the symmetry-restored wavefunction is multireference in character since it is composed of a combination of determinants with different spin configurations. This can be seen as a configuration interaction expansion among a set of non-orthogonal determinants of size equal to the number of basis functions. The energy of an SUHF wavefunction is,
\begin{center}\begin{equation}
E = \frac{\langle \Phi| P^{s\dagger}_{mm} H P^s_{mm}|\Phi\rangle}{\langle \Phi| P^{s\dagger}_{mm}P^s_{mm}|\Phi\rangle} =  \frac{\langle \Phi| H P^s_{mm}|\Phi\rangle}{\langle \Phi| P^s_{mm}|\Phi\rangle}.
\end{equation}\end{center}
Note here that projection operators are Hermitian and idempotent, $P=P^\dagger=P^2$, and $[H,P]=0$ since we project symmetries of the Hamiltonian. As the UHF reference determinant is already an eigenstate of $S_z$, the integration of $\alpha$ and $\gamma$ can be eliminated, \cite{ref10}
\begin{center}\begin{equation}
 P^s_{mm} = \frac{2s+1}{2}\int^\pi_0\mathrm d\beta\sin\beta d^s_{mm}(\beta)e^{\mathrm i\beta S_y},
\end{equation}\end{center}
where $d^s_{mm}(\beta)=\langle s;m|e^{\mathrm i\beta S_y}|s;m\rangle$ is the Wigner d-matrix. This operator projects a UHF wavefunction with $S_z$ quantum number $m$ to a new wavefunction with total spin quantum number $s$. We use this projection to generate a reference state and add correlations with the Gutzwiller factor to further improve the results. For the following discussion, we will drop the subscripts of the projection operator with the understanding that we are preserving the $S_z$ eigenvalue of the reference determinant. \footnote{Details on the evaluation of projected states are discussed in the supplemental material.~\cite{ref11}}

\subsection{The Gutzwiller Similarity Transformation}

Additional short-range screening effects are added through a similarity transformation of the Hamiltonian as mentioned above. In a previous paper,~\cite{ref8} we introduced a transformation generated by a Hilbert space Jastrow operator constructed out of two-body products of on-site number operators, 
\begin{center}\begin{equation}
n_{i\sigma}= c^\dagger_{i\sigma}c_{i\sigma}.
\end{equation}\end{center}
The operators $ c^\dagger_{i\sigma}$ and $ c_{i\sigma}$ are on-site fermion creation and annihilation operators, and $\sigma = \uparrow,\downarrow$ is the spin index. The original transformation was generated by all two-body combinations of the occupancy operators including double-occupancy and products of spin and density operators. Here we build on the previous work while focusing only on the local term in the transformation that yields short-range correlations,
\begin{center}\begin{equation}
J = \sum_i \alpha_i n_{i\uparrow}n_{i\downarrow},
\end{equation}\end{center}
where the local weights $\alpha_i$ are real parameters to be optimized. This operator, applied to a wavefunction as a correlation factor, has the form of a locally weighted double-occupancy screening operator or Gutzwiller factor. This is equivalent to the more common form of the Gutzwiller factor with local weights, \cite{ref5}
\begin{center}\begin{equation}
e^J = \prod_i\Big(1-(1-g_i)n_{i\uparrow}n_{i\downarrow}\Big), \quad g_i = e^{\alpha_i}
\end{equation}\end{center}
 While containing fewer terms than a more general two-body Jastrow, the local correlations are found to be most important for short range interactions.~\cite{ref8} This factor is applied as a similarity transformation to the Hamiltonian as in Eq. (\ref{eq3}). The transformation is evaluated by resumming the Baker-Campbell-Hausdorff expansion into a local one-body rotation,~\cite{ref11}
\begin{center}\begin{equation}
e^{-J} c^\dagger_{i\uparrow}e^{J} = c^\dagger_{i\uparrow}e^{-\alpha_in_{i\downarrow}} \label{eq10}.
\end{equation}\end{center}
We refer to this as the Gutzwiller similarity transformation (GST). Each fermion operator acquires a weight based on the on-site density of the opposite spin electrons. Due to the simple nature of the operator it becomes useful to rewrite the exponential form of the local transformation using the idempotency of the number operator ($n_{i\sigma}^2 = n_{i\sigma}$),~\cite{ref11}
\begin{center}\begin{equation}
e^{-\alpha_in_{i\sigma}} = 1+(e^{-\alpha_i}-1) n_{i\sigma} \label{eq12}.
\end{equation}\end{center}
This result makes the transformation of general operators straightforward. By applying the transformation to each of the creation and annihilation operators and using Eq. (\ref{eq10},\ref{eq12}) for a one-body operator we obtain,
\begin{center}\begin{equation}
e^{-J} c^\dagger_{i\uparrow} c_{j\uparrow}e^{J} = (1+\xi^-_i n_{i\downarrow}) c^\dagger_{i\uparrow} c_{j\uparrow}(1+\xi^+_j n_{j\downarrow}),
\end{equation}\end{center}
where $\xi^+_i=e^{\alpha_i}-1$ and $\xi^-_i = e^{-\alpha_i}-1$.

We will now apply this transformation to the nearest-neighbor, repulsive Hubbard Hamiltonian,
\begin{center}\begin{equation}
 H = -t\sum_{\langle ij\rangle}( c^\dagger_{i\uparrow} c_{j\uparrow} +  c^\dagger_{i\downarrow} c_{j\downarrow}) + U\sum_i  n_{i\uparrow} n_{i\downarrow}.
\end{equation}\end{center}
This is a highly studied, non-trivial system containing only local interactions but long-range correlations. In this Hamiltonian, $\langle ij\rangle$ represents nearest-neighboring sites, $t$ is the energy of a particle hopping between neighboring sites, and $U$ is the interaction between two particles on the same site. While a mean-field wavefunction screened by a Gutzwiller factor is not the correct solution to the nearest-neighbor Hubbard model, this ansatz and other similar wavefunctions have been used to study this Hamiltonian and prove to be good approximations.~\cite{ref3,ref5,ref7,ref8}

Applying the similarity transformation to the Hubbard Hamiltonian results in a new, non-Hermitian, three-body Hamiltonian $\overline H = e^{-J} He^{J}$,~\cite{ref5}
\newpage
\begin{eqnarray}
{\overline H} = &- t\sum_{\langle ij\rangle} \Big((1+\xi^-_i n_{i\downarrow}) c^\dagger_{i\uparrow} c_{j\uparrow} (1+\xi^+_j n_{j\downarrow}) \nonumber\\
 &+(1+\xi^-_i n_{i\uparrow}) c^\dagger_{i\downarrow} c_{j\downarrow}(1+\xi^+_j n_{j\uparrow})\Big) \nonumber\\
&+ U\sum_i  n_{i\uparrow} n_{i\downarrow}.  
\end{eqnarray}
By writing the Hamiltonian as a three body operator, we can maintain mean-field computational cost when evaluating it over Slater determinant states. This new Hamiltonian has the same spectrum as the original, but we have introduced local screening effects with a set of parameters $\alpha_i$ that must be optimized. Given a trial wavefunction, we aim to select the parameters in such a way that an eigenstate of the transformed Hamiltonian is best approximated by the test state.

\subsection{The Calculation Scheme}
We now describe the procedure to calculate the parameters in the ansatz. We refer to the following equations, before any symmetry projection operators are applied, as the unrestricted Gutzwiller similarity transformation (UGST). In order to perform calculations, we construct a system of equations to solve for the amplitudes $\alpha_i$ and the reference Slater determinant. Since the transformed Hamiltonian is non-Hermitian, we do not expect the left and right eigenstates to be the same. We therefore use a biorthogonal ansatz for increased flexibility in the optimization,
\begin{center}\begin{equation}
E=\langle\Phi_L| {\overline H}|\Phi_R\rangle, \label{eq15}
\end{equation}\end{center}
where $|\Phi_L\rangle$ and $|\Phi_R\rangle$ are UHF Slater determinants with different single particle bases and intermediate normalization $\langle\Phi_L|\Phi_R\rangle = 1$. We find this has some advantages aside from providing a more general ansatz than a single determinant. The results can be more accurate, particularly for doped lattices, and the stability and convergence rate of the reference optimization process is significantly improved.

As $\overline H$ is non-Hermitian, calculated energies are not an upper bound to the ground state. Therefore, we cannot optimize the parameters in the Ritz variational sense. Instead, the degrees of freedom are selected by requiring the energy to be stationary under a set of constraints. This is regularly done by multiplying the constraints by a set of Lagrange multipliers and adding them to the energy.~\cite{ref1,ref3} The resulting Lagrangian is
\begin{center}\begin{equation}
L = E + \sum_i z_iR_i. \label{eq16}
\end{equation}\end{center}
$E$ is the energy (\ref{eq15}), $R_i$ is a set of constraints we impose, and $z_i$ are the corresponding Lagrange multipliers. The additional constraints are required for a better general optimization scheme and the subsequent calculation of many relevant observables. Any quantity that commutes with the Gutzwiller factor such as spin, density, and double occupancy would otherwise be treated purely on the mean-field level. The constraints are constructed by projection of Schr\"{o}dinger's equation
\begin{center}\begin{equation}
\overline H|\Phi_R\rangle = E|\Phi_R\rangle,
\end{equation}\end{center}
 into a set of states, $\{\langle \Phi_L|n_{i\uparrow}n_{i\downarrow}\}$, defined by the components of $J$,
\begin{center}\begin{equation}
\langle \Phi_L|n_{i\uparrow}n_{i\downarrow}\overline H|\Phi_R\rangle = E\langle \Phi_L|n_{i\uparrow}n_{i\downarrow}|\Phi_R\rangle.
\end{equation}\end{center}
We require $|\Phi_R\rangle$ to be an eigenstate within the set of states spanned by the component operators of the Gutzwiller correlator. The constraints are then defined as
\begin{center}\begin{equation}
R_i = \langle\Phi_L| n_{i\uparrow} n_{i\downarrow}(\overline H - E)|\Phi_R\rangle\label{eq18}.
\end{equation}\end{center}
These conditions can equivalently be defined by requiring the energy variance to be zero within the projected subspace.~\cite{ref5} This Lagrangian is similar in form to those used in other similarity transformation methods such as coupled cluster, where the equations are projected into a set of excited determinants.~\cite{ref1} While we could in principle use any set of states, we select this set in order to consider fluctuations most relevant to the on-site Gutzwiller correlation factors we use.

The optimization conditions of all the degrees of freedom are now defined by taking derivatives of $L$,
\begin{eqnarray}
\frac{\partial L}{\partial z_i} = R_i = 0, \forall i, \\
\frac{\partial L}{\partial \alpha_i} = 0, \forall i.
\end{eqnarray}
This system of equations is solved for the parameters $\alpha_i$ and $z_i$. In order to optimize the right and left reference determinants, we use Hartree-Fock self-consistent field equations. A generalized Fock matrix is constructed as a derivative of the Lagrangian with respect to the one-particle transition density $\rho$,~\cite{ref12}
\begin{center}\begin{equation}
F_{i\sigma,j\sigma'} = \frac{\partial L}{\partial \rho_{j\sigma',i\sigma}}, \label{eq21}
\end{equation}\end{center}
where,
\begin{eqnarray}
\rho_{i\sigma,j\sigma'} = \langle \Phi_L| c^\dagger_{j\sigma'} c_{i\sigma}|\Phi_R\rangle, \\
\rho = C^o_R C^{o\dagger}_L. \nonumber
\end{eqnarray}
We use the normalization condition,
\begin{eqnarray}
C^{o\dagger}_LC^o_R = I. \label{eq23}
\end{eqnarray}
$C^o_L$ and $C^o_R$ are $M\times N_o$ matrices containing the occupied orbital coefficients of the left and right Slater determinants respectively, where $M$ is the number of spin orbitals and $N_o$ is the number of occupied states. As the left and right states are constructed out of the left and right eigenvectors of the Fock matrix at each iteration of the optimization process, the overlap matrix (\ref{eq23}) is diagonal by construction. The reference determinants are calculated with standard self-consistent Hartree-Fock iterations until $F$ and $\rho$ share common left and right eigenbases indicating we have reached a stationary point.~\cite{ref12} Both the amplitude equations and the Generalized Fock matrix have low computational cost, scaling as $\mathcal O(M^2)$ for the Hubbard Hamiltonian after construction of the transition density. \footnote{Explicit expressions for the energy and residuals are provided in the supplemental material. \cite{ref11}}

Once we have the reference determinants, we use the projection operators to restore symmetry of the wavefunction and further improve the results. We refer to this as spin-projected UGST (SUGST).  In principle, the reference optimization above can be done in the presence of the projection operators, but we find that this does not significantly change the results. In addition, the cost and difficulty of converging the equations is dramatically increased. As a result, we choose to leave the reference determinants unchanged at this point and solve for a new set of amplitudes $\alpha^s_i$ and $z^s_i$ in the presence of the projection. The expression for the energy and amplitude equations for the projected wavefunctions are the same as before,
\begin{eqnarray}
E^s = \langle P^s_L|\overline H^s | P^s_R \rangle, \label{eq24} \\
R^s_i = \langle P^s_L| n_{i\uparrow}n_{i\downarrow}(\overline H^s - E^s) | P^s_R\rangle, \label{eq25}\\
L^s = E^s + \sum_i z^s_iR^s_i,\label{eq26}
\end{eqnarray}
where $\overline H^s$ is the transformed Hamiltonian evaluated with $\alpha^s_i$, and
\begin{eqnarray}
|P^s_R\rangle = \frac{P^s|\Phi_R\rangle}{\sqrt{\langle\Phi_L|P^s|\Phi_R\rangle}},\\
\langle P^s_L| = \frac{\langle\Phi_L|P^s}{\sqrt{\langle\Phi_L|P^s|\Phi_R\rangle}}\nonumber.
\end{eqnarray}
As before, the values of $\alpha^s_i$ and $z^s_i$ are calculated by requiring,
\begin{eqnarray}
R^s_i = 0,\forall i, \\
\frac{\partial L^s}{\partial\alpha^s_i} = 0, \forall i.
\end{eqnarray}

Expectation values of observables other than the energy are evaluated with linear response. Response densities are calculated through derivatives of $L^s$ with respect to elements of the Hamiltonian. The one and two-particle response densities are, \cite{ref3} 
\begin{eqnarray}
\Gamma = & &\langle P^s_L|\overline\Gamma |P^s_R\rangle\label{eq30}  \\
&+& \sum_q z_q \langle P^s_L| n_{q\uparrow}n_{q\downarrow} \Big(\overline\Gamma - \langle P^s_L|\overline\Gamma|P^s_R\rangle\Big)|P^s_R\rangle \nonumber
\end{eqnarray}
where $\overline{\Gamma}=e^{-J}a^\dagger_{j\sigma'}a_{i\sigma}e^J$ for the one-particle density and $\overline{\Gamma}= e^{-J}a^\dagger_{i\sigma}a^\dagger_{j\sigma'}a_{l\gamma'}a_{k\gamma}e^J$ for the two-particle density. Operators such as spin, density, and double occupancy that commute with the transformation are still modified  by the correlations when calculating their expectation value.
It is important to note here that individual elements of the spin density do not commute with the spin operators. Unless the calculated property commutes with $S^2$ and $S_z$, the projections must be applied to the left and right determinants and a full spin projection operator must be used.

\vspace{2ex}

\noindent The iterative procedure for optimization of the reference determinants and correlation amplitudes is as follows:
\begin{itemize}
\item [\textbf{1.}] Make an initial guess for $C_L$ and $C_R$.
\item [\textbf{2.}] Solve $R_i=0$ for the amplitudes $\alpha_i$.
\item [\textbf{3.}] Solve $\frac{\partial L}{\partial\alpha_i} = 0$ for the response amplitudes $z_i$.
\item [\textbf{4.}] Construct and diagonalize the Fock matrix to build a new set of coefficients $C_L$ and $C_R$.
\item [\textbf{5.}] Iterate 2-4 until the equations converge and the Hartree-Fock condition $[F,\rho]=0$ is satisfied
\item [\textbf{6.}] Solve $R^s_i=0$ for the amplitudes $\alpha^s_i$.
\item [\textbf{7.}] Solve $\frac{\partial L^s}{\partial\alpha^s_i} = 0$ for the response amplitudes $z^s_i$.
\end{itemize}
The optimized parameters are then used to calculate energy and other properties with Eq. (\ref{eq15},\ref{eq24},\ref{eq30}).

\vspace{1ex}

\section{Results and discussions}

We present benchmark calculations on Hubbard systems and compare the results to available accurate data.  All calculations are performed on lattices with periodic boundary conditions, the spin state $s=m=0$, and energies reported in units of $t$. We also compare to unrestricted coupled-cluster singles and doubles (UCCSD) where correlations are introduced with a Hamiltonian similarity transformation consisting of all single and double excitation operators that preserve $S_z$ symmetry evaluated with the UHF determinant.~\cite{ref1}  Tables are provided in the supplemental material for direct comparison.~\cite{ref11}

\begin{figure*}[ht!]
 \centering
 \includegraphics[width= \textwidth]{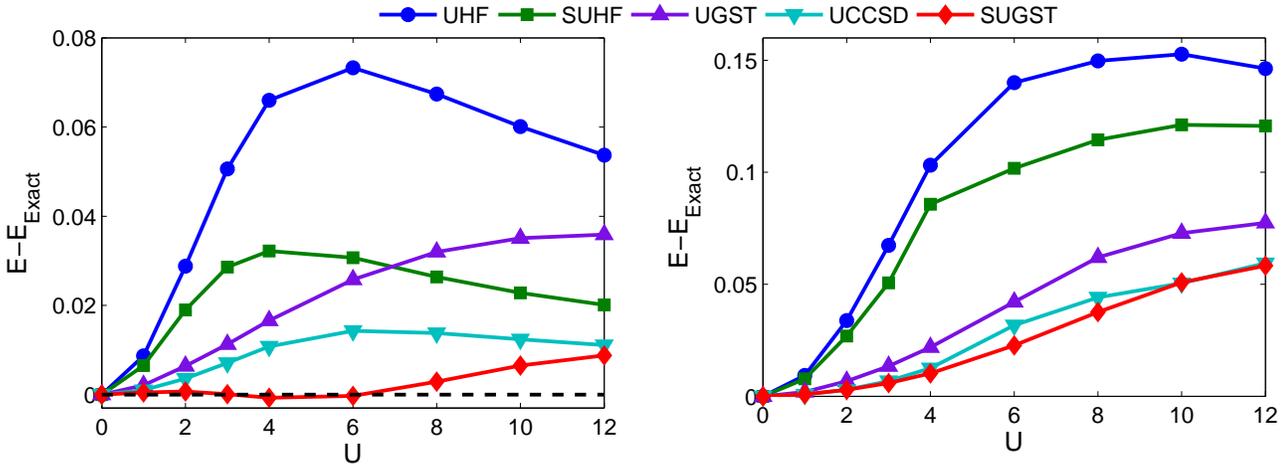}
 \caption{\label{fig1} Error in the energy per site for $4\times4$ square Hubbard lattices with 16 particles (left) and 14 particles (right) compared to exact diagonalization from Ref.~\cite{ref13}.}
\end{figure*}

In Figure~\ref{fig1}, we compare results for some $4\times 4$ square lattices where exact energies are available. It is clear that SUHF and UGST have significant improvements over the UHF reference energy. When the two methods are combined in SUGST, the result is cumulative and we capture more of the correlation energy, typically more than UCCSD. The SUGST correlation is less than the sum of the SUHF and UGST correlation energy indicating there may be some overlap in the correlation energy recovered. The quality of the results diminishes slightly for the doped systems as spin projection of a GHF state is likely better suited, but the energies remain similar to UCCSD.

\begin{figure}[h]
\centering
 \includegraphics[width=.5\textwidth]{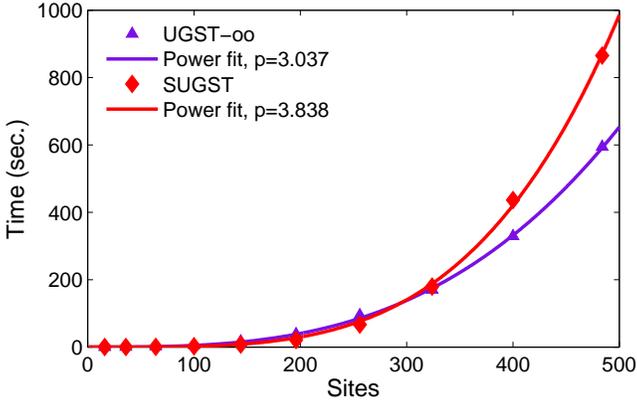}
 \caption{\label{fig2} Time required for the UGST orbital optimization and SUGST amplitude optimization with integration grid size equal to $\sqrt{N_{sites}}$ on half-filled Hubbard square lattices with $U=4$}
\end{figure}

A significant advantage of SUGST is the low cost of the calculations. As discussed above, UGST scales as $\mathcal O(M^3)$ in the number of sites which matches the observed times very closely (Figure~\ref{fig2}). The SUGST calculation formally scales as $\mathcal O(M^3N)$ where $N$ is the size of the integration grid, which is slightly lower than the observed rate. This may vary for different systems as the convergence rates can change. The low scaling means we can easily perform calculations on large systems with relatively little computational effort.

\begin{figure*}[ht!]
 \centering
 \includegraphics[width=\textwidth]{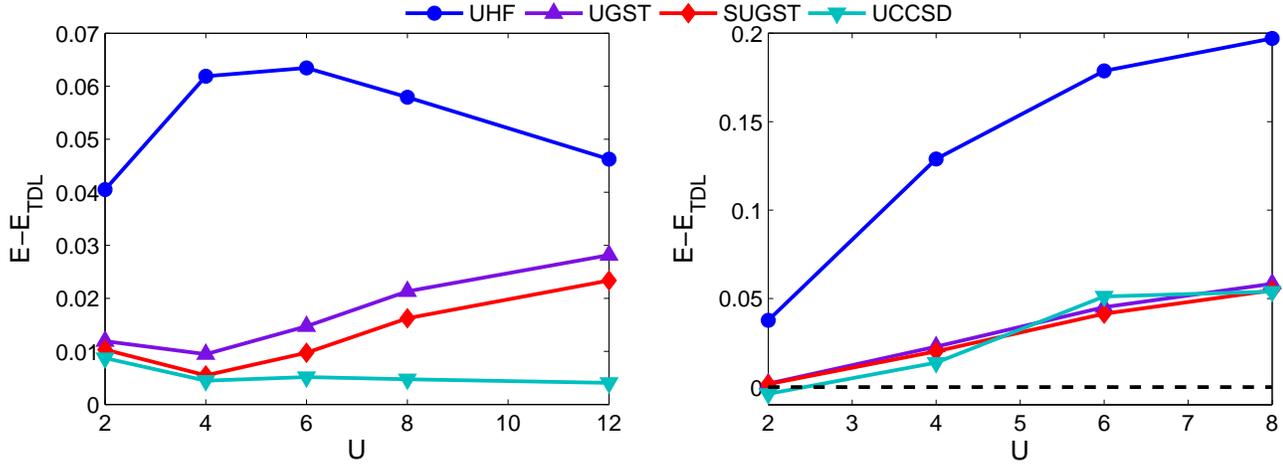}
 \caption{\label{fig3} Energy errors and finite-size effects per site for $10\times10$ square Hubbard lattices with $\langle n\rangle =1$ (left) and $\langle n \rangle = 0.8$ (right) compared to UCCSD and averages of high quality thermodynamic limit calculations ($E_{TDL}$) from Refs.~\cite{ref14,ref15} as accurate results for finite systems are limited. }
\end{figure*}

We now apply the method to a set of larger Hubbard lattices at varying values of $U$ and compare the results to UCCSD (Figure~\ref{fig3}). There are some finite-size effects apparent for smaller values of $U$ as the reference energies are borrowed from calculations for infinite systems.~\cite{ref15} 
Again we see significant improvement over mean-field when the transformation is applied and evaluated with the projected wavefunctions.
 Unlike the $4\times 4$ case, there is some reduction in accuracy for larger $U$ at half-filling. For the smaller systems, much of the correlation energy in this case was recovered through the projected wavefunction, and comparatively less was recovered by UGST than in the doped cases. In the larger systems, there is significantly less correlation energy per-site recovered with the projection, hence the larger errors.

\begin{figure}[h]
 \centering
 \includegraphics[width=.5\textwidth]{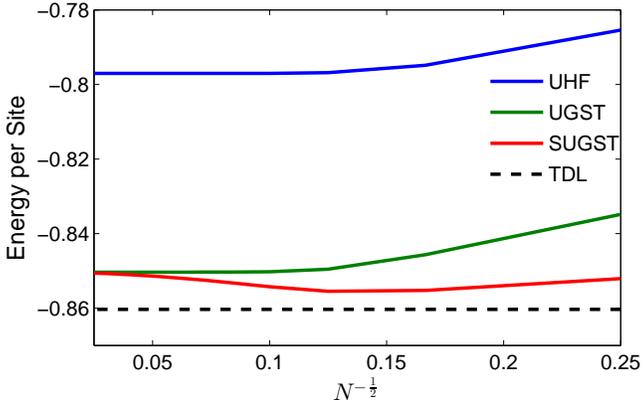}
 \caption{\label{fig4} Energies per site for the half-filled Hubbard model approaching the thermodynamic limit for $N$-site square lattices at $U=4$, with an average of high quality results for an infinite system (TDL) from Ref.~\cite{ref15}.}
\end{figure}

We can see the reason for the reduction in quality as we increase the lattice size by observing the effects in the thermodynamic limit. As the computational cost of SUGST is low, it is a simple matter to perform calculations on very large systems. In Figure~\ref{fig4} we show the size effects on the energy per site of square Hubbard lattices. It is clear that SUGST suffers from the same lack of size extensivity observed in PHF. \cite{ref10} UGST converges to a thermodynamic limit as UHF does, but the additional correlation energy from the projection decreases as the system size increases once the thermodynamic limit is reached and eventually returns to the UGST energy per particle. There is a size intensive term in projection that yields a finite constant to be added to the infinite energy of an infinite system.~\cite{ref10}

\begin{figure*}[t!]
 \centering
 \includegraphics[width=\textwidth]{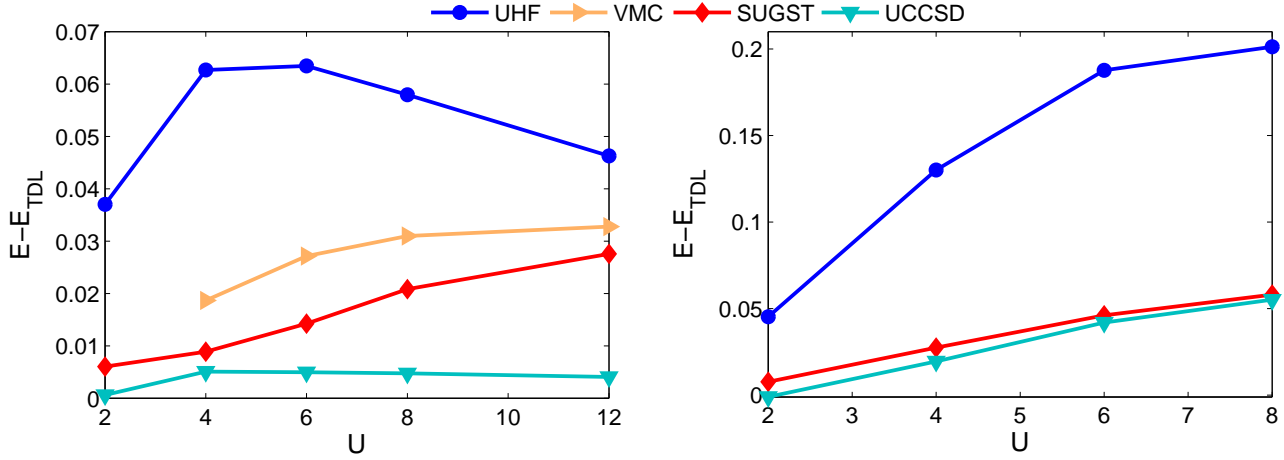}
 \caption{\label{fig5} Energy errors per site for $30\times30$ square Hubbard lattices with $\langle n\rangle =1$ (left) and $\langle n \rangle = 0.8$ (right) compared to extrapolated variational Monte Carlo (VMC) from Ref.~\cite{ref7} as well as extrapolated UCCSD and averages of high quality thermodynamic limit calculations ($E_{TDL}$) from Ref.~\cite{ref15}.}
\end{figure*}

\begin{figure}[h]
 \centering
 \includegraphics[width=.5\textwidth]{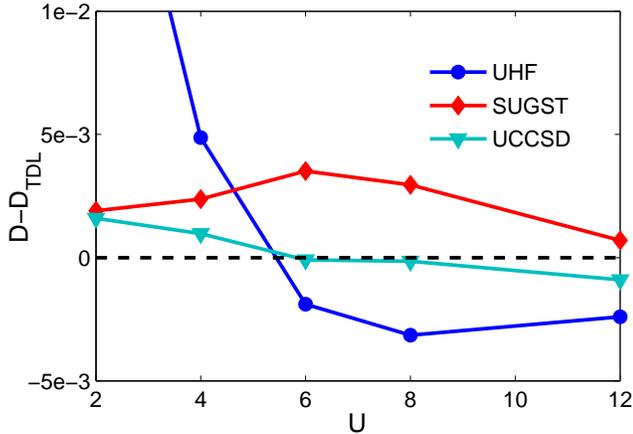}
 \caption{\label{fig6} Double occupancy errors per site for $30\times30$ square Hubbard lattices with $\langle n\rangle =1$  compared to UCCSD and average of high quality thermodynamic limit calculations ($D_{TDL}$) from Ref.~\cite{ref15}.}
\end{figure}

If we now compare results in the thermodynamic limit for different values of 
$U$, we can see the previously observed behavior is maintained for large systems (Figure~\ref{fig5}). As the SUHF wavefunction brings effectively no correlation energy per site for such large systems, we again see a reduction in accuracy for the strongly correlated case at half-filling with large $U$. We still find reasonable accuracy for $U=4$ and the doped cases with results very close to the largest UCCSD lattices available. We also compare the double occupancy of the large lattices calculated with the response densities. (Figure~\ref{fig6}). While SUGST slightly overestimates the double occupancy, the error does not vary widely as UHF does.

We also compare to the results taken from the literature~\cite{ref7} using variational Monte Carlo calculations with a Gutzwiller factor and an antiferromagnetic, mean-field reference. As the symmetry projection provides negligible improvement when approaching the thermodynamic limit, and all the local correlation factors equal a constant value for half-filling, this is a reasonable comparison with the variational solution of the wavefunction. In Figure~\ref{fig5}, we can see that both the variational and projective methods have similar errors. We can further directly compare the wavefunctions by looking at the correlation amplitudes and antiferromagnetic order parameter in Table~\ref{table1}. While we do not get exactly the same energies and parameters, the results are very similar. We do not'
 expect the results to be identical as the Gutzwiller wavefunction is not an exact solution. However, the similarity in the results indicates that we have made a good approximation to the variational solution without the need for Monte Carlo sampling.
 
 \Table{\label{table1} Gutzwiller correlation factors ($g$) and antiferromagnetic order parameters ($M$) from SUGST on a $30\times30$, half-filled lattice and variational Monte Carlo (VMC) extrapolated to the thermodynamic limit.}
\br
\ns
$U$ & 4 & 6 & 8 & 12 \\
\mr
$g_{VMC}{}^a$  & 0.65 & 0.55 & 0.50 & 0.40 \\
$g_{SUGST}$    & 0.6167 & 0.5205 & 0.4679 & 0.4147 \\\mr
$M_{VMC}{}^{a,b}$ & 0.58(2)  & 0.77(1) & 0.86(1) & 0.92(1) \\
$M_{SUGST}$   & 0.5851 & 0.7422 & 0.8352 & 0.9256 \\
\br
\end{tabular}
\item [] ${}^{\rm a}$ Results taken from  Ref.~\cite{ref7}.
\item [] ${}^{\rm b}$ Uncertainty for the last digit is given in parentheses.
\end{indented}
\end{table}

\begin{figure}[t]
 \centering
 \includegraphics[width=.5\textwidth]{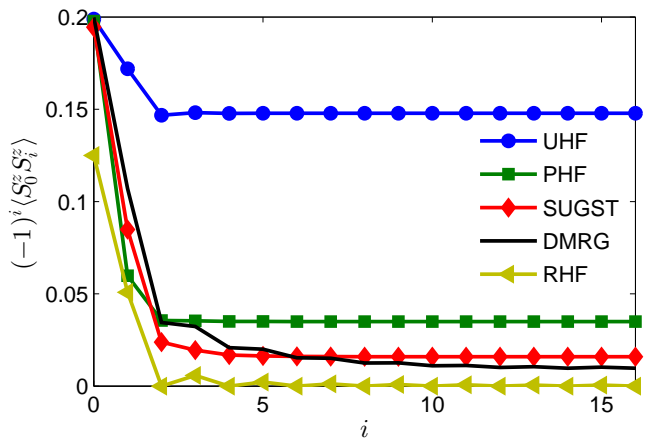}
 \caption{\label{fig7} Spin-spin correlation function with alternating sign for a 30-site Hubbard ring with $U=4$ compared to the exact density matrix renormalization group theory (DMRG) result.~\cite{ref16,ref17}}
\end{figure}

As the Gutzwiller factor only includes on-site terms, 
it provides significant improvement for short-range quantities such as the energy and double occupancy discussed above. In this method, longer-range correlations are left entirely to the reference wavefunction. We see that the spin-spin correlation function (Figure \ref{fig7}) quickly decays to a constant value and does not capture the correct long range decay of the exact correlation. There is significant improvement over the Hartree-Fock and projected results as the correlations are not severely over or underestimated in the medium range. The short range interactions are effectively screened, allowing the reference wavefunction to approximate the long-range effects more accurately within the limitations of the ansatz. As shown in previous work, long-range effects can be correctly calculated if the Jastrow factor used for the transformation contains long-range terms.

\section{Conclusions}

We have presented a similarity transformed model generated by the Guzwiller factor that produces high quality energies for weak and intermediate correlation when evaluated with an optimized, biorthogonal reference. We can easily evaluate the Hamiltonian with a projected reference further improving the results in the strongly correlated regimes for smaller systems. In addition, the calculated double occupancies are consistently close to the best available data, and the errors do not vary greatly for different interaction strengths as in the mean-field calculations. We have also shown that our results are similar to variational Monte Carlo calculations with similar wavefunctions, indicating our method is a good approximation to the variational solution. The projected wavefunction corrects much of the error in the strongly correlated cases, but the additional energy from projection suffers from lack of size extensivity. It could be effectively used in smaller lattices as an impurity solver for embedding methods such as density matrix embedding theory.~\cite{ref18}

The results are comparable to and sometimes better than UCCSD, a much more costly method scaling at $\mathcal O(M^5)$ for the Hubbard model versus $\mathcal O(M^3)$ in UGST. Some of the current shortfalls could be addressed by evaluating UGST with more advanced wavefunctions and Jastrow factors. Long-range terms can be included in the transformation to improve the description of the correlation functions.~\cite{ref8} Projected GHF and multireference projected wavefunctions are likely candidates that build on the current results as they better address doped and strongly correlated systems on larger lattices.~\cite{ref19,ref20} They would also provide a framework to calculate excited states in order to explore the energy spectrum and are a subject of future study.

\vspace{2ex}

\noindent \textbf{Acknowledgements} - The authors thank Dr. Thomas M. Henderson for assistance with coupled cluster calculations. This work was supported by the National Science Foundation (CHE-1462434) and the Welch Foundation (C-0036). This research used resources of the National Energy Research Scientific Computing Center, a DOE Office of Science User Facility supported by the Office of Science of the U.S. Department of Energy under Contract No. DE-AC02-05CH11231.

\end{document}